\documentclass[sigconf, review=false]{acmart}
\usepackage{subcaption}
\usepackage{booktabs} % For formal tables
\setcopyright{rightsretained}
\acmDOI{10.48550/arXiv.2410.22566}

\acmConference[ICVGIP'24]{14th Indian Conference on Computer Vision, Graphics and Image Processing}{December 2024}{Bangalore, India}
\begin{document}
% \title{ICVGIP 2021 LaTeX Template}
\title{Deep Priors for Video Quality Prediction} 

% \titlenote{Produces the permission block, and

\author{Siddharath Narayan Shakya \\ \texttt{s23048@students.iitmandi.ac.in} 
  % \and Parimala Kancharla \\ \texttt{parimala@example.com}
  }

\affiliation{
  \institution{Indian Institute of Technology Mandi}
  % \streetaddress{XYZ}
  \city{Mandi}
  \state{Himachal Pradesh}
  \country{India}
  \postcode{175005}
}

\author{ Parimala Kancharla \\ \texttt{parimala@iitmandi.ac.in}}

\affiliation{
  \institution{Indian Institute of Technology Mandi}
  % \streetaddress{XYZ}
  \city{Mandi}
  \state{Himachal Pradesh}
  \country{India}
  \postcode{175005}
}

% \author{Parimala Kancharla}
% \affiliation{
%   \institution{Indian Institute of Technology Mandi}
%   % \streetaddress{XYZ}
%   \city{Mandi}
%   \state{Himachal Pradesh}
%   \country{India}
%   \postcode{175005}
% }

% \author{Siddharath Shakya,~\IEEEmembership{Student Member,~IEEE,},Parimala Kancharla,~\IEEEmembership{Member,~IEEE,}
% }

\renewcommand{\shortauthors}{}

\begin{abstract}
   In this work, we designed a completely blind video quality assessment algorithm using the deep video prior. This work mainly explores the utility of deep video prior in estimating the visual quality of the video. In our work, we have used a single distorted video and a reference video pair to learn the deep video prior. At inference time, the learned deep prior is used to restore the original videos from the distorted videos. The ability of learned deep video prior to restore the original video from the distorted video is measured to quantify distortion in the video.  Our hypothesis is that the learned deep video prior fails in restoring the highly distorted videos. The restoring ability of deep video prior is proportional to the distortion present in the video. Therefore, we propose to use the distance between the distorted video and the restored video as the perceptual quality of the video.  Our algorithm is trained using a single video pair and it does not need any labelled data. We show that our proposed algorithm  outperforms the existing unsupervised video quality assessment algorithms in terms of LCC and SROCC on a synthetically distorted video quality assessment dataset.
   
\end{abstract}




\maketitle

\section{Introduction}
Video content and the scale of video consumption has been increasing due to high speed internet and mobile devices.  Good objective video quality assessment (VQA) algorithms are required to efficiently manage this content.
Objective VQA algorithms are classified into three categories: i) Full-Reference (FR) ii) Reduced-Reference (RR) and iii) No-Reference (NR) methods. FR-VQA algorithms require distorted video and also the corresponding reference video. RR-VQA algorithms require partial knowledge of the reference video while NR-VQA algorithms do not need any knowledge of the reference video for predicting the quality of the test video. We can classify the NR-VQA algorithms into sub categories: i) Supervised methods, which use subjective quality scores as labels while designing the algorithm.
ii) Unsupervised methods (or blind methods), which do not use any subjective quality scores (labelled data) while designing the algorithm.
\\
The main drawback of the Supervised NR-VQA algorithms is generalizability across the distortion types and  also across the datasets. So it is important to design a universal video quality predictor, which is not biased towards datasets. There are few unsupervised NR-VQA algorithms \cite{mittal2012making, mittal2015completely, tu2020comparative} in the literature. The unsupervised video quality algorithms VIIDEO \cite{mittal2015completely} and NIQE \cite{mittal2012making} build on the statistical priors learned from natural videos and natural images respectively. The low level statistics of images and videos are modelled using probabilistic models. The Natural Image Quality Evaluator (NIQE) \cite{mittal2012making} assumes that all the natural images posses a unique statistical signature. The mean subtracted contrast normalised coefficients (MSCN) of the natural images follows a Gaussian distribution \cite{ruderman1994statistics}. However, the distribution of MSCN coefficients of the distorted images do not follow a Gaussian distribution. The statistical prior model describing the set of natural images is estimated. In order to get the visual quality of the distorted image, NIQE \cite{mittal2012making} computes the distance between the statistical model of the distorted image and the statistical prior estimated from the corpus of natural images. In order to get the video level quality, mean of  NIQE scores of all the frames is computed. The completely blind video oracle \cite{mittal2015completely} models the MSCN coefficients of frame differences using a statistical model. The statistical model estimated from the set of natural videos is used as a prior. To estimate a test video's quality, it measures how far is the test video's statistical model from the reference statistical model. 
\\
Statistical priors have been used thus far in designing the video quality assessment algorithms. Recently, frameworks such as Deep Image Prior (DIP) \cite{ulyanov2018deep} and Deep Decoder (DD) \cite{heckel2018deep} have been proposed to capture image priors. These frameworks claim that the structure of a convolutional neural network (CNN) is sufficient to capture the low level image  statistics. The CNN structure itself acts as a prior. The Deep Video Prior (DVP) model \cite{lei2020blind} is proposed to impose temporal consistency while learning the model on a pair of original and processed frames. In general, natural videos are smooth and temporally consistent.  In the literature, temporal consistency is enforced explicitly computing optical flow or motion vectors while solving video tasks. The DVP model \cite{lei2020blind} claims that the feature maps of a CNN on two similar patches are expected to be similar at the early stage of training. The objective function of DVP is proposed to ensure that CNN feature maps of the predicted frames and the ground truth frames are similar. This kind of objective function is sufficient to impose temporal consistency. This captures the smoothness present in natural videos. So, the model learned using this objective function acts as the prior describing the natural videos.  All these deep prior models \cite{ulyanov2018deep},\cite{heckel2018deep} and \cite{lei2020blind} show that excellent performance can be achieved by using only test data.
Unlike most previous methods that enforce temporal consistency with optical flow, this method shows that the temporal consistency can be achieved by training a convolutional network on a video with the Deep Video Prior. We have used this deep prior model to design an blind video quality algorithm. These models are performing well in many computer vision tasks like super resolution and denoising and enhancement.
\section{Proposed Method}
% \begin{comment}
\begin{figure}[htbp]
\includegraphics[width=3.5in]{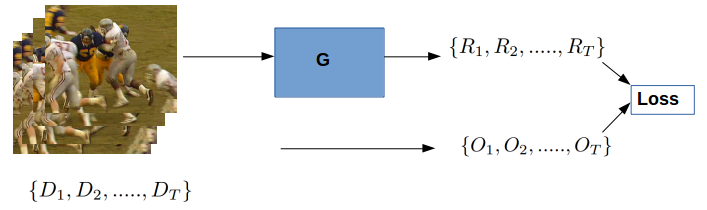}
\caption{Training the Deep Video Prior}
\label{fig:block_diagram}
\end{figure}
% \end{comment}
\vspace{-0.2cm}
Given the efficacy of the DVP \cite{lei2020blind} in video colorization and video enhancement tasks, we propose to use this model in designing an unsupervised video quality assessment algorithm. 
A Fully Convolutional Network $(G)$ is used to learn the distortion operator as shown in fig. \ref{fig:block_diagram}.
The main objective of the model $G$ is to restore the original frames from the given distorted frames. The temporal consistency is enforced through loss function.  
$\{D_{t}\}_{t=1}^{T}$ represents the distorted video and $\{O_{t}\}_{t=1}^{T}$ represents the original video. The distorted frames $\{D_{t}\}_{t=1}^{T}$ are passed through the model $G$ to get the restored frames $\{ R_{t} = G(D_{t})\}_{t=1}^{T}$. We train the model such that the restored frames $\{R_{t}\}_{t=1}^{T}$ look like the original frames $\{O_{t}\}_{t=1}^{T}$.  We have used the objective function described in equation \ref{eqn:block_diagram2} to update the weights $\theta$ of the network $G$. The loss function $L_{\text{perceptual}}$ (\ref{eqn:block_diagram2}) computes the $L_{1}$ norm between the features of restored frame $R_{t}$ and the original frame $O_{t}$ at each layer of the network $G$. $F_{k}$ indicates the CNN features of the $k^{th}$ layer. We have used a Fully Convolutional Network based on  VGG-19 architecture for model $G$. 
A single video pair (the distorted video $\{D_{t}\}_{t=1}^{T}$ and the original video $\{O_{t}\}_{t=1}^{T}$ is used to train the model $G$. The model $G$ is initialized randomly and trained for 10 epochs.
% \vspace{-0.5cm}
\begin{equation}
     L_{\text{perceptual}}(G(D_{t}),O_{t})   = \min_{\theta \epsilon \mathcal{G}} \sum_{k=1}^{L}||F_{k}(G(D_{t})-F_{k}(O_{t})||_{1} 
    \label{eqn:block_diagram2}
\end{equation}
% \vspace{-0.7cm}
\subsection{Quality Prediction}
Generally, the natural videos are inherently smooth and temporally consistent. Therefore, quantifying the temporal inconsistency gives us the perceptual quality score of a video
The learned model $G$ is further used to predict the quality on the test videos.
We pass the test videos (distorted videos) through the model $G$. 
We predict the visual quality of the distorted video by measuring how well the deep video prior is restoring the original frames from the distorted frames. 
We build on the fact that highly distorted videos are not restored well using the deep video prior. The distance between distorted frames and restored frames is used as quality of the video. The higher distortion in the video leads to artifacts in restored frames as well. So the model $G$ fails in reconstructing the  original frames well. We claim that the distance between restored frames and distorted frames is proportional to visual quality of the video. The equation \ref{eqn:psnr_equation} describes the computation of the video quality score.
% \vspace{-0.4cm}
\begin{equation}
\text{Quality Score} = \frac{1}{T} \sum _{t=1}^{T} \text{log}({\text{PSNR}}(R_{t},D_{t})).
\label{eqn:psnr_equation}
\end{equation}
% \vspace{-0.4cm}

The mean of the log of peak signal to noise ratio (PSNR) between restored frames and distorted frames is used as he quality score of the video.  $T$ is the number of frames in the given test video.
\section{Results and Discussion}
We used the ECVQ dataset \cite{6338461} to show the improved performance of our quality prediction algorithm. It has 90 videos of resolution $352 \times 288$. The Linear Correlation Coefficient (LCC) and Spearman Rank Order Correlation Coefficient (SROCC) scores are used to evaluate the video quality assessment algorithms.  The LCC and SROCC are computed between the subjective scores of the videos and predicted video quality scores. The higher values of LCC and SROCC indicates that the proposed algorithm is correlating well with the human perpetual quality. We compared our proposed method with the completely blind algorithms. Our results are reported in the table \ref{tab:konvid_scores}. 
Our proposed method is outperforming both the methods, which build on the statistical priors. 

\begin{table}[htbp]
\centering

    \begin{tabular}{|c|c|c|}
    
    \hline
        \textbf{Method} & \textbf{LCC} & \textbf{SROCC}\\
         \hline
         NIQE \cite{mittal2012making}  & 0.4960 & 0.4469 \\
         %\hline 
        % V- blinds in paper & & \\
         \hline 
        VIIDEO \cite{mittal2015completely} & 0.28 & 0.15\\
        
  \hline                    

\textbf{Proposed Method}   & \textbf{0.5089}& \textbf{0.5209}  \\
        
\hline
     
    \end{tabular}
    \caption{Results on ECVQ Dataset \cite{hosu2017konstanz} for completely blind video quality assessment}
    \label{tab:konvid_scores}
\end{table}
% \vspace{-0.7cm}
\section{Conclusion}
% \vspace{-0.2cm}
Through this preliminary attempt, we demonstrate  that  DVP can be used in designing a video quality predictor. To the best of our knowledge, this is the first work that successfully applies the deep learning based prior to the video quality prediction. Our method uses only a single reference and test video pair for training the DVP. The models \cite{lei2020blind, heckel2018deep, ulyanov2018deep} have been trained using single test sample are shown to be performing well in other computer vision tasks like denoising and colorization etc. However, these frameworks have not been applied yet to quality assessment tasks. We believe that this opens up a new direction for video or image quality assessment research to use the deep video and image priors. 
As future work, we plan to build on these preliminary results by using more advanced image priors on a frame wise basis. We would like to extend this approach further to design an efficient unsupervised stand-alone approach for video quality prediction.

% {\small
% \bibliographystyle{ieee_fullname}
% \bibliography{egbib}
% }

\bibliographystyle{ACM-Reference-Format}
\bibliography{ICVGIP-Latex-Template}

% \appendix


% The appendix gets added after the references.


\end{document}